\documentclass[preprint,prd,aps,showpacs,showkeys,nofootinbib]{revtex4}
\usepackage{graphicx}
\begin{document}
\title{Search for charged lepton flavor violation of vector mesons in BLMSSM model}
\author{Xing-Xing Dong$^{1}$\footnote{dxx$\_$0304@163.com},
Shu-Min Zhao$^{1}$\footnote{zhaosm@hbu.edu.cn},
 Jing-Jing Feng$^{1}$,
 Guo-Zhu Ning$^{1}$\footnote{nkninggz@163.com},\\
Jian-Bin Chen$^{2}$\footnote{j.b.chen1007@gmail.com},
 Hai-Bin Zhang$^{1}$\footnote{hbzhang@hbu.edu.cn},
Tai-Fu Feng$^{1}$\footnote{fengtf@hbu.edu.cn}}
\affiliation{$^1$ Department of Physics and Technology, Hebei University, Baoding 071002, China\\
$^2$ Department of Physics, Taiyuan University of Technology, Taiyuan 030024, China}
\begin{abstract}
We analyze the charged lepton flavor violating (CLFV) decays of vector mesons $V\rightarrow l_i^{\pm}l_j^{\mp}$ with $V\in\{\phi, J/\Psi, \Upsilon, \rho^0, \omega \}$ in BLMSSM model. This new model is introduced as an supersymmetric extension of Standard Model (SM), where local gauged baryon number B and lepton number L are considered. The numerical results indicate BLMSSM model can produce significant contributions to such two-body CLFV decays. And the branching ratios for these CLFV processes can easily reach the present experimental upper bounds. Therefore, searching for CLFV processes of vector mesons may be an effective channels to study new physics.
\end{abstract}


\maketitle

\section{Introduction}
The neutrino oscillation experiments have convinced that neutrinos possess tiny masses and mix with each other\cite{neutrino1,neutrino2,neutrino3,neutrino4,neutrino5,neutrino6,neutrino7,neutrino8,neutrino9}. This phenomenon shows that charged lepton flavor violating  (CLFV) process \cite{CLFV1,CLFV2,CLFV3} is strongly suppressed in the Standard Model (SM). Therefore, CLFV processes are considered as an evidence to study new physics beyond the SM. Physicists do more research on the CLFV decays of vector mesons in various SM extension, such as grand unified models~\cite{grand unified1,grand unified2}, supersymmetric models with and without R-party\cite{supersymmetric models1,supersymmetric models2,supersymmetric models3}, left-right symmetry models\cite{symmetry models1,symmetry models2} and Z¡¯ models\cite{Z' model1,Z' model2} etc. In our previous work, we investigate these CLFV processes in the framework of MSSM with type I seesaw mechanism\cite{MSSM type I1,MSSM type I2}, and some of the theoretical evaluations on CLFV processes fit better with the experimental upper bounds, such as $J/\Psi(\Upsilon)\rightarrow l_i^{\pm}l_j^{\mp}$ with $l_i, l_j\in\{e, \mu, \tau \}$. However, the predictions on processes $\phi(\rho^0,\omega)\rightarrow e^+\mu^-$ in MSSM with type I seesaw mechanism are around $10^{-20}$, which are far below the present experimental upper bounds.

Current experiments naturally suggest to search for two-body decays of vector mesons in the $e^{\pm}\mu^{\mp}$ final state. Considering the experimental constraints on the processes $l_j\to 3l_i$ and $l_j\to l_i+\gamma$, many experts have studied the CLFV processes involving vector mesons $V\rightarrow l_i^{\pm}l_j^{\mp}$ with $V\in\{\phi, J/\Psi, \Upsilon, \rho^0, \omega \}$ and $l_i, l_j\in\{e, \mu, \tau \}$\cite{constraints1,constraints2}. Likewise, the constraints from  $\mu-e$ conversion are also taken into account on these processes\cite{constraints3,constraints4,constraints5}. Currently, PDG\cite{PDG} give an independent experimental upper limits on the two-body decays of vector mesons, which are shown as
\begin{eqnarray}
&&Br(\phi\rightarrow e^+\mu^-) \leq 2.0 \times 10^{-6},\;\;\;\;\; Br(J/\Psi\rightarrow e^+\mu^-) \leq 1.6 \times 10^{-7},
\nonumber\\&&Br(J/\Psi\rightarrow e^+\tau^-) \leq 8.3 \times 10^{-6},\; Br(J/\Psi\rightarrow \mu^+\tau^-) \leq 2.0 \times 10^{-6},
\nonumber\\&&Br(\Upsilon\rightarrow \mu^+\tau^-) \leq 6.0 \times 10^{-6}.
\end{eqnarray}

As an extension of the minimal supersymmetric standard model (MSSM)\cite{supersymmetric models1,supersymmetric models2,MSSM1,MSSM2}, gauged baryon number (B) and lepton number (L) are added in the BLMSSM~\cite{BLMSSM1,BLMSSM2,BLMSSM3,BLMSSM4}. Compared with the MSSM possessing R-party conservation, there are new parameters and new contributions to CLFV processes. We introduce the local gauged $B$ to explain the matter-antimatter asymmetry in the universe. Lepton number $L$ is expected to be broken spontaneously at the TeV scale. Furthermore, right-handed neutrinos are considered to explain the neutrino oscillation experiments. In our work, the CLFV processes of vector mesons ($\phi(J/\Psi, \Upsilon, \rho^0, \omega)\rightarrow e^+\mu^-$  and $J/\Psi(\Upsilon)\rightarrow e^+\tau^-(\mu^+\tau^-)$) are investigated within the BLMSSM. Let us assume that a vector meson $V_i$ couples to $e^{\pm}\mu^{\mp}$, and the corresponding effective Lagrangian can be written  as\cite{constraints2,constraints3,constraints4,constraints5}:
\begin{eqnarray}
{\cal L}_{{eff}}=V_i^{\nu}(\xi_V^M\bar{\mu}\gamma_{\nu}e+\xi_A^M\bar{\mu}\gamma_{\nu}\gamma_5e+h.c.),
\end{eqnarray}
where $V_i$ is a quark-antiquark bound state like $\phi, J/\Psi, \Upsilon, \rho^0, \omega$. $\xi_{V,A}^M$ represent the effective couplings of the vector meson $V_i$ to the CLFV lepton currents.

This work is organized as follows: In Sec.2, we describe the BLMSSM model briefly, including the corresponding interaction lagrangian, needed mass matrices and couplings. As an example, we derive the analytic results of amplitudes for diagrams in Section 3. In Sec.4, we give out the corresponding parameters and numerical results. And the conclusion is shown in Section 5. The superfields in the BLMSSM are presented in Appendix A.
\section{BLMSSM}
In this work, we study a supersymmetric model where baryon $(B)$ and lepton $(L)$ numbers are local gauge symmetries. This model is defined as BLMSSM and the corresponding local gauge group is $SU(3)_C\otimes{SU(2)_L}\otimes{U(1)_Y}\otimes{U(1)_B}\otimes{U(1)_L}$~\cite{BLMSSM1,group1,group2}. In the BLMSSM, the local $B$ and $L$ are spontaneously broken at the TeV scale. In order to cancel the $B$ and $L$ anomalies, a vector-like family is needed, which are $\hat{Q}_4, \hat{U}_4^c, \hat{D}_4^c, \hat{L}_4, \hat{E}_4^c, \hat{N}_4^c$ and $\hat{Q}_5^c, \hat{U}_5, \hat{D}_5, \hat{L}_5^c, \hat{E}_5, \hat{N}_5 $. And the corresponding superfields presented in BLMSSM are shown in TABLE I of Appendix A. To break baryon number and lepton number spontaneously, the model introduces Higgs superfields $\hat{\Phi}_B$, $\hat{\varphi}_B$ and $\hat{\Phi}_L$ $\hat{\varphi}_L$ respectively. After these Higgs superfields acquiring nonzero vacuum expectation values (VEVs), the exotic quarks and leptons obtain masses. Furthermore, the introduction of superfields $\hat{X}$ and $\hat{X}'$ is to make exotic quarks avoid stability. Actually, the lightest one can be a dark matter candidate.

The superpotential of BLMSSM is written as
\begin{eqnarray}
&&{\cal W}_{BLMSSM}={\cal W}_{MSSM}+{\cal W}_{B}+{\cal W}_{L}+{\cal W}_{X}\;,
\end{eqnarray}
here ${\cal W}_{MSSM}$ represents the superpotential of the MSSM, and the concrete forms of superpotentials ${\cal W}_{B}$, ${\cal W}_{L}$ and ${\cal W}_{X}$ are given as

\begin{eqnarray}
&&{\cal W}_{B}=\lambda_{Q}\hat{Q}_{4}\hat{Q}_{5}^c\hat{\Phi}_{B}+\lambda_{U}\hat{U}_{4}^c\hat{U}_{5}
\hat{\varphi}_{B}+\lambda_{D}\hat{D}_{4}^c\hat{D}_{5}\hat{\varphi}_{B}+\mu_{B}\hat{\Phi}_{B}\hat{\varphi}_{B}
\nonumber\\&&\hspace{1.2cm}+Y_{{u_4}}\hat{Q}_{4}\hat{H}_{u}\hat{U}_{4}^c+Y_{{d_4}}\hat{Q}_{4}\hat{H}_{d}\hat{D}_{4}^c
+Y_{{u_5}}\hat{Q}_{5}^c\hat{H}_{d}\hat{U}_{5}+Y_{{d_5}}\hat{Q}_{5}^c\hat{H}_{u}\hat{D}_{5},
\nonumber\\&&{\cal W}_{L}=Y_{{e_4}}\hat{L}_{4}\hat{H}_{d}\hat{E}_{4}^c+Y_{{\nu_4}}\hat{L}_{4}\hat{H}_{u}\hat{N}_{4}^c
+Y_{{e_5}}\hat{L}_{5}^c\hat{H}_{u}\hat{E}_{5}+Y_{{\nu_5}}\hat{L}_{5}^c\hat{H}_{d}\hat{N}_{5}
\nonumber\\&&\hspace{1.2cm}+Y_{\nu}\hat{L}\hat{H}_{u}\hat{N}^c+\lambda_{{N^c}}\hat{N}^c\hat{N}^c\hat{\varphi}_{L}
+\mu_{L}\hat{\Phi}_{L}\hat{\varphi}_{L},
\nonumber\\&&{\cal W}_{X}=\lambda_1\hat{Q}\hat{Q}_{5}^c\hat{X}+\lambda_2\hat{U}^c\hat{U}_{5}\hat{X}^\prime
+\lambda_3\hat{D}^c\hat{D}_{5}\hat{X}^\prime+\mu_{X}\hat{X}\hat{X}^\prime.
\end{eqnarray}
The soft breaking terms $\mathcal{L}_{soft}$ in the BLMSSM can be found in our previous work~\cite{soft1,soft2,soft3}.

In the BLMSSM, the $SU(2)_L$ doublets $H_{u}$, $H_{d}$ and singlets $\Phi_{B}$, $\varphi_{B}$, $\Phi_{L}$, $\varphi_{L}$ obtain the nonzero VEVs $\upsilon_{u}$, $\upsilon_{d}$ and $\upsilon_{B}$, $\overline{\upsilon}_{B}$, $\upsilon_{L}$, $\overline{\upsilon}_{L}$ respectively, then the local gauge symmetry $SU(2)_L\otimes{U(1)_Y}\otimes{U(1)_B}\otimes{U(1)_L}$ breaks down to the electromagnetic symmetry $U(1)_e$.
\begin{eqnarray}
&&H_u=\left({\begin{array}{*{20}{c}}
H_u^+  \\
\frac{1}{\sqrt 2}(\upsilon_u+H_u^0+iP_u^0)  \\
\end{array}}
\right), H_d=\left({\begin{array}{*{20}{c}}
\frac{1}{\sqrt 2}(\upsilon_d+H_d^0+iP_d^0)  \\
H_d^-  \\
\end{array}}
\right),\nonumber\\
&&\Phi_B=\frac{1}{\sqrt{2}} (\upsilon_B+\Phi_B^0+iP_B^0),\;\;\;\;\; \varphi_B=\frac{1}{\sqrt{2}} (\overline{\upsilon}_B+\varphi_B^0+i\overline{P}_B^0),\nonumber\\
&&\Phi_L=\frac{1}{\sqrt{2}}(\upsilon_L+\Phi_L^0+iP_L^0),\;\;\;\;\;\;\varphi_L=\frac{1}{\sqrt{2}}
(\overline{\upsilon}_L+\varphi_L^0+i\overline{P}_L^0).
\end{eqnarray}

In this model, we introduce the superfields $\hat{N}^c$, so three neutrinos obtain tiny masses through the see-saw mechanism. In the basis $(\psi_{\nu_L^I}, \psi_{N_R^{cI}})$, the mass matrix of neutrinos is deduced after the symmetry breaking
\begin{eqnarray}
&&Z_{N_{\nu}}^\top\left(\begin{array}{cc}
  0&\frac{v_u}{\sqrt{2}}(Y_{\nu})^{IJ} \\
   \frac{v_u}{\sqrt{2}}(Y^{T}_{\nu})^{IJ}  & \frac{\bar{v}_L}{\sqrt{2}}(\lambda_{N^c})^{IJ}
    \end{array}\right)Z_{N_{\nu}}=diag(m_{\nu^{\alpha}}), \alpha=1\cdot\cdot\cdot6,I,J=1,2,3,
\nonumber\\&& \psi_{{\nu_L^I}}=Z_{{N_{\nu}}}^{I\alpha}k_{N_\alpha}^0,\;\;\;\;
\psi_{N_R^{cI}}=Z_{{N_{\nu}}}^{(I+3)\alpha}k_{N_\alpha}^0,\;\;\;\;
\chi_{N_\alpha}^0=\left(\begin{array}{c}
   k_{N_\alpha}^0\\  \bar{k}_{N_\alpha}^0
    \end{array}\right).
\end{eqnarray}
here, $\chi_{N_\alpha}^0$ represent the mass eigenstates of neutrino fields mixed by left-handed and right-handed neutrinos.

Similarly, with the introduced superfields $\hat{N}^c$, we can also obtain the mass squared matrix of sneutrinos in the base $\tilde{n}^T=(\tilde{\nu},\tilde{N}^c)$. And this matrix is more complicated than that in MSSM.
\begin{eqnarray}
&&\left(\begin{array}{cc}
  {\cal M}^2_{\tilde{n}}(\tilde{\nu}_{I}^*\tilde{\nu}_{J})&{\cal M}^2_{\tilde{n}}(\tilde{\nu}_I\tilde{N}_J^c) \\
   ({\cal M}^2_{\tilde{n}}(\tilde{\nu}_I\tilde{N}_J^c))^{\dag} & {\cal M}^2_{\tilde{n}}(\tilde{N}_I^{c*}\tilde{N}_J^c)
    \end{array}\right),
\end{eqnarray}
where,
\begin{eqnarray}
  && {\cal M}^2_{\tilde{n}}(\tilde{\nu}_{I}^*\tilde{\nu}_{J})=\frac{g_1^2+g_2^2}{8}(v_d^2-v_u^2)\delta_{IJ}+g_L^2(\overline{v}^2_L-v^2_L)\delta_{IJ}
   +\frac{v_u^2}{2}(Y^\dag_{\nu}Y_\nu)_{IJ}+(m^2_{\tilde{L}})_{IJ},\nonumber\\&&
   {\cal M}^2_{\tilde{n}}(\tilde{N}_I^{c*}\tilde{N}_J^c)=-g_L^2(\overline{v}^2_L-v^2_L)\delta_{IJ}
   +\frac{v_u^2}{2}(Y^\dag_{\nu}Y_\nu)_{IJ}+2\overline{v}^2_L(\lambda_{N^c}^\dag\lambda_{N^c})_{IJ}\nonumber\\&&
   \hspace{2.9cm}+(m^2_{\tilde{N}^c})_{IJ}+\mu_L\frac{v_L}{\sqrt{2}}(\lambda_{N^c})_{IJ}
   -\frac{\overline{v}_L}{\sqrt{2}}(A_{N^c})_{IJ}(\lambda_{N^c})_{IJ},\nonumber\\&&
   {\cal M}^2_{\tilde{n}}(\tilde{\nu}_I\tilde{N}_J^c)=\mu^*\frac{v_d}{\sqrt{2}}(Y_{\nu})_{IJ}-v_u\overline{v}_L(Y_{\nu}^\dag\lambda_{N^c})_{IJ}
   +\frac{v_u}{\sqrt{2}}(A_{N})_{IJ}(Y_\nu)_{IJ}.
\end{eqnarray}
Through matrix $Z_{\nu}$, the mass matrix can be diagonalized.

With the new gaugino $\lambda_L$ and the superpartners of  $SU(2)_L$ singlets $\Phi_{L}$, $\varphi_{L}$ mixing together, the mass matrix of lepton neutralinos $\chi_L^0$ is produced. In the base $(i\lambda_L, \psi_{\Phi_L}, \psi_{\varphi_L})$, we can diagonalize the mass matrix $\chi_L^0$ by $Z_{N_L}$.
\begin{equation}
\mathcal{L}_{\chi_L^0}=\frac{1}{2}(i\lambda_L,\psi_{\Phi_L},\psi_{\varphi_L})\left(     \begin{array}{ccc}
  2M_L &2v_Lg_L &-2\bar{v}_Lg_L\\
   2v_Lg_L & 0 &-\mu_L\\-2\bar{v}_Lg_L&-\mu_L &0
    \end{array}\right)  \left( \begin{array}{c}
 i\lambda_L \\ \psi_{\Phi_L}\\\psi_{\varphi_L}
    \end{array}\right)+h.c.
\end{equation}

From the contributions of the superpotential and the soft breaking terms, the corrected form for the slepton mass squared matrix reads as
\begin{eqnarray}
&&\left(\begin{array}{cc}
  (\mathcal{M}^2_L)_{LL}&(\mathcal{M}^2_L)_{LR} \\
   (\mathcal{M}^2_L)_{LR}^{\dag} & (\mathcal{M}^2_L)_{RR}
    \end{array}\right),
\end{eqnarray}
where,
\begin{eqnarray}
 &&(\mathcal{M}^2_L)_{LL}=\frac{(g_1^2-g_2^2)(v_d^2-v_u^2)}{8}\delta_{IJ} +g_L^2(\bar{v}_L^2-v_L^2)\delta_{IJ}
 +m_{l^I}^2\delta_{IJ}+(m^2_{\tilde{L}})_{IJ},\nonumber\\&&
 (\mathcal{M}^2_L)_{LR}=\frac{\mu^*v_u}{\sqrt{2}}(Y_l)_{IJ}-\frac{v_u}{\sqrt{2}}(A'_l)_{IJ}+\frac{v_d}{\sqrt{2}}(A_l)_{IJ},
 \nonumber\\&& (\mathcal{M}^2_L)_{RR}=\frac{g_1^2(v_u^2-v_d^2)}{4}\delta_{IJ}-g_L^2(\bar{v}_L^2-v_L^2)\delta_{IJ}
 +m_{l^I}^2\delta_{IJ}+(m^2_{\tilde{R}})_{IJ}.
\end{eqnarray}
Then the mass matrix can be rotated to the mass eigenstates by the unitary matrix $Z_{\tilde{L}}$.

In the BLMSSM, the introduced superfields $\hat{N}^c$ lead to corrections for the couplings existed in MSSM. We deduce some corrected couplings, such as W-lepton-neutrino and Z-neutrino-neutrino couplings, which are shown as:
\begin{eqnarray}
&&\mathcal{L}_{WL\nu}=-\frac{e}{\sqrt{2}s_W}W_{\mu}^+\sum_{I=1}^3\sum_{\alpha=1}^6Z_{N_{\nu}}^{I\alpha*}\bar{\chi}_{N_{\alpha}}^0\gamma^{\mu}P_Le^I,
\nonumber\\&&\mathcal{L}_{Z\nu\nu}=-\frac{e}{2s_Wc_W}Z_{\mu}\sum_{I=1}^3\sum_{\alpha,\beta=1}^6Z_{N_{\nu}}^{I\alpha*}Z_{N_{\nu}}^{I\beta}\bar{\chi}_{N_{\alpha}}^0\gamma^{\mu}P_L\chi_{N_{\beta}}^0,
\end{eqnarray}
where $P_L=\frac{1-\gamma_5}{2}$ and $P_R=\frac{1+\gamma_5}{2}$. We define $s_W=\sin\theta_W$, $c_W=\cos\theta_W$, and $\theta_W$ is the Weinberg angle.

The Z-sneutrino-sneutrino coupling is deduced as:
\begin{eqnarray}
&&\mathcal{L}_{Z\tilde{\nu}\tilde{\nu}}=-\frac{e}{2s_Wc_W}Z_{\mu}\sum_{I=1}^3\sum_{i,j=1}^6Z_{\nu}^{Ii*}Z_{\nu}^{Ij}\tilde{\nu}^{i*}i(\overrightarrow{\partial}^{\mu}
-\overleftarrow{\partial}^{\mu})\tilde{\nu}^j.
\end{eqnarray}

We also obtain the chargino-lepton-sneutrino coupling:
\begin{eqnarray}
&&\mathcal{L}_{\chi^{\pm}L\tilde{\nu}}=-\sum_{I=1}^3\sum_{i=1}^6\sum_{j=1}^2\bar{\chi}^-_j
\Big(Y_l^{I} Z_-^{2j*}Z_{\nu}^{Ii*}P_R+
[\frac{e}{s_W}Z_+^{1j}Z_{\nu}^{Ii*}
\nonumber\\&&\hspace{1.8cm}+Y_\nu^{Ii}Z_+^{2j}Z_{\nu}^{(I+3)i*}]P_L
\Big)e^I\tilde{\nu}^{i*}+h.c.
\end{eqnarray}

Considering the interactions of gauge and matter multiplets
$ig\sqrt{2}T^a_{ij}(\lambda^a\psi_jA_i^*-\bar{\lambda}^a\bar{\psi}_iA_j)$,
we deduce a new coupling for lepton-slepton-lepton neutralino. The corresponding form for this coupling is written as
\begin{eqnarray}
&&\mathcal{L}_{l\chi_L^0\tilde{L}}=\sqrt{2}g_L\bar{\chi}_{L_j}^0\Big(Z_{N_L}^{1j}Z_{\tilde{L}}^{Ii}P_L
-Z_{N_L}^{1j*}Z_{\tilde{L}}^{(I+3)i}P_R\Big)l^I\tilde{L}_i^++h.c.
\end{eqnarray}
\section{The amplitudes for CLFV decays of vector mesons}
In the BLMSSM, the CLFV decays of vector mesons $V\rightarrow l_i^{\pm}l_j^{\mp}$ with $V\in\{\phi, J/\Psi, \Upsilon, \rho^0, \omega \}$ and $l_i, l_j\in\{e, \mu, \tau \}$ are studied. We know that meson is consist of quark and anti-quark. Meson $\phi$ is made up of $s\bar{s}$; $J/\Psi$ is constituted of $c\bar{c}$; $\Upsilon$ is composed of $b\bar{b}$; $\rho^0$ is comprised of $\frac{1}{\sqrt{2}}(u\bar{u}-d\bar{d})$ and $\omega$ is consist of $\frac{1}{\sqrt{2}}(u\bar{u}+d\bar{d})$.  We depict the relevant Feynman diagrams contributing to these processes in FIG.1 and FIG.2.

\begin{figure}[h]
\centering
\includegraphics[width=15cm]{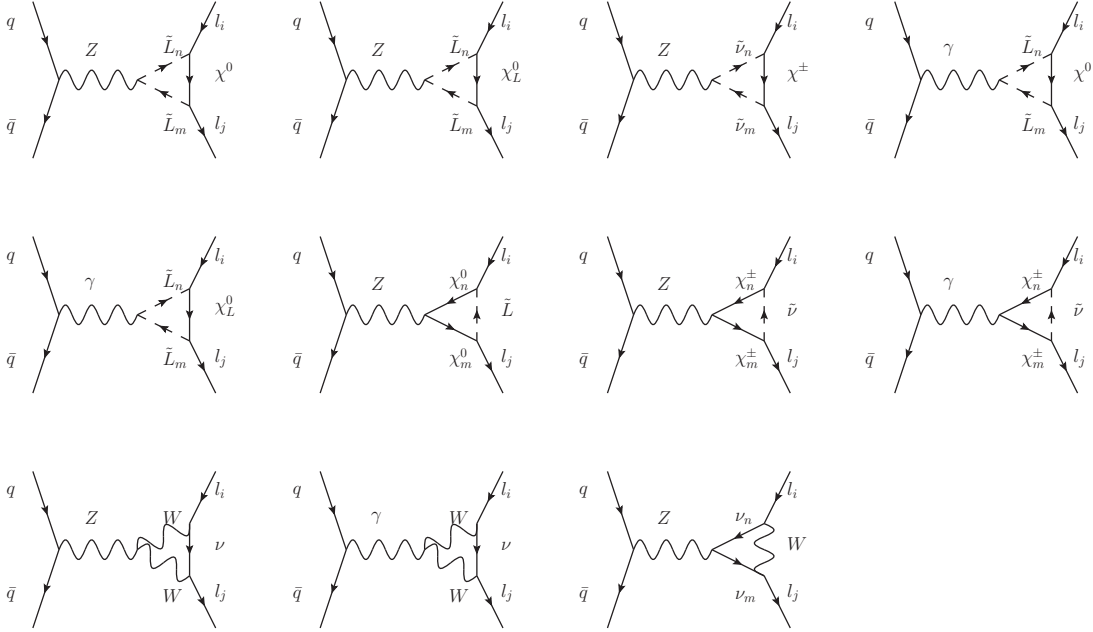}\\
\caption{The penguin-type diagrams for processes $V\rightarrow l_i^{\pm}l_j^{\mp}$, with $q$ representing $u$, $c$, $d$, $s$, $b$.} \label{fig1}
\end{figure}

\begin{figure}[h]
\centering
\includegraphics[width=8cm]{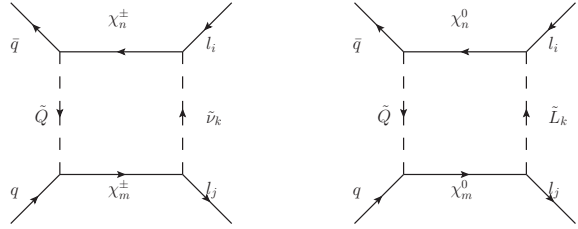}\\
\caption{The box-type diagrams for processes $V\rightarrow l_i^{\pm}l_j^{\mp}$.} \label{fig2}
\end{figure}
In the quark picture, mesons are composed of a quark and an anti-quark. QCD has the property of "quark confinement", so the traditional way to the perturbative calculation can not work. At quark-gluon level, the complicated calculation of loop integrations is governed by the non-perturbative QCD effects. However, completely reliable way to these non-perturbative QCD effects is lacking at present. Therefore, as a powerful phenomenological model, a sum rule for light-cone wavefunction\cite{light-cone1,light-cone2,light-cone3,light-cone4,light-cone5} is adopted, which is widely used in the theoretical research of particle physics and nuclear physics.

To obtain the decay amplitude of process involving a vector meson, one needs to calculate the matrix elements of gauge invariant nonlocal operators\cite{light-cone4,light-cone5}
\begin{eqnarray}
&&\langle0|\bar{q}(y)\Gamma[x,y]q(x)|V\rangle,
\end{eqnarray}
where $\Gamma[x,y]$ is a generic Dirac matrix structure, x and y represent the coordinates of quark and anti-quark. The leading-twist distribution amplitude of vector meson $V$ can be defined by the correlator~\cite{light-cone4,light-cone5}:
\begin{eqnarray}
&&\langle0|\bar{q}_{\alpha}(y)q_{\beta}(x)|V(p)\rangle=\frac{\delta_{ij}}{4N_c}\int_0^1due^{-i(upy+\bar{u}px)}
[f_Vm_V{/\!\!\!\varepsilon}_VV_{||}(u)
\nonumber\\&&\hspace{4.2cm}+\frac{i}{2}\sigma^{\mu'\nu'}f_V^{\mathrm{\top}}(\varepsilon_{V\mu'}p_{\nu'}
-\varepsilon_{V\nu'}p_{\mu'})V_{\bot}(u)]_{\beta\alpha},
\end{eqnarray}
where $m_V$ and $\varepsilon_V$ are respectively the mass and polarization vector of the vector meson, $f_V$ and $f_V^{\top}$ are the meson decay constants, $V_{||}(u)$ and $V_{\bot}(u)$ represent the leading-twist distribution functions corresponding to the longitudinally and transversely polarized mesons. The momentum $p$ satisfies $p^2=m_V^2$, which shows that meson momentum is on-shell. The integration variable $u$ stands for the momentum fraction carried by the quark, and $\bar{u}\equiv1-u$ corresponds to the momentum fraction of anti-quark. In our calculation, we make $V_{||}=V_{\bot}=V(u)=6u(1-u)$, the reason is that the meson amplitudes are similar to their asymptotic form\cite{asymptotic form}. We take the number of colors $N_c=3$.

Taking one penguin diagram in FIG.1 and one box diagram in FIG.2 as examples, we show how to calculate the CLFV decays for vector mesons in BLMSSM. In the frame of center of mass, the amplitude of decay $\phi\rightarrow e^+\mu^-$ in FIG.1(1) can be shown at hadron level:
\begin{eqnarray}
&&{\cal A}_{\phi1}=\langle e^+\mu^-|\bar{u}_{\mu}(p_4)
\int\frac{d^Dk}{(2\pi)^D}\frac{-g_{\mu\nu}A_2({/\!\!\!k}+m_{\chi_k^0})A_1A_3(2k-p_3+p_4)^{\mu}}
{[k^2-m_{\chi_k^0}^2][(k-p_3)^2-m_{\tilde{L}_i}^2][(k+p_4)^2-m_{\tilde{L}_j}^2]}v_e(p_3)
\nonumber\\&&\hspace{1.2cm}\times\bar{v}_s(p_2)\frac{A_4}{[(p_1+p_2)^2-m_Z^2]}u_s(p_1)|\phi(p)\rangle
\end{eqnarray}
Here,
\begin{eqnarray}
&&A_1=[\frac{e}{\sqrt{2}s_Wc_W}Z_{\tilde{L}}^{Jj}(Z_N^{1k}s_W+Z_N^{2k}c_W)+Y_l^JZ_{\tilde{L}}^{(J+3)j}Z_N^{3k}]P_L
\nonumber\\&&\hspace{1.2cm}+[\frac{-\sqrt{2}e}{c_W}Z_{\tilde{L}}^{(J+3)j}Z_N^{1k*}+Y_l^JZ_{\tilde{L}}^{Jj}Z_N^{3k*}]P_R, \nonumber\\&&A_2=(A_1)_{j\rightarrow i,J\rightarrow I}^*,
\nonumber\\&&A_3=\frac{e}{2s_Wc_W}(Z_{\tilde{L}}^{Ki}Z_{\tilde{L}}^{Kj*}-2s_W^2\delta^{ij}),
\nonumber\\&&A_4=\frac{e}{2s_Wc_W}\gamma^{\nu}(P_L-\frac{2}{3}s_W^2)
\end{eqnarray}

$\sum|n\rangle\langle n|\equiv1$ and the vacuum state $|0\rangle\langle 0|$ is the dominate item. In the amplitude ${\cal A}_{\phi1}$, $|0\rangle\langle 0|\simeq 1$ is considered to put between wavefunctions $v_e(p_3)$ and $\bar{v}_s(p_2)$. Applying the Eq.(17), $\langle0|\bar{v}_s(p_2)A_4u_s(p_1)|\phi(p)\rangle= i\frac{ef_{\phi}m_{\phi}(3-4s_W^2)}{12N_cs_Wc_W}\varepsilon^{\mu*}(p)$. In order to simplify amplitude ${\cal A}_{\phi1}$, the general one-loop tensor N-point integrals\cite{N-point1,N-point2} are used
\begin{eqnarray}
T_{\mu_1\cdot\cdot\cdot\mu_p}^N(p_1,\cdot\cdot\cdot,p_{N-1},m_0,\cdot\cdot\cdot,m_{N-1})
=\frac{(2\pi\mu)^{4-D}}{i\pi^2}d^Dq\frac{q_{\mu_1}\cdot\cdot\cdot q_{\mu_p}}{D_0D_1\cdot\cdot\cdot D_p},
\end{eqnarray}
here, the denominator factors are $D_0=q^2-m_0^2+i\varepsilon, D_i=(q+p_i)^2-m_i^2+i\varepsilon, i=1,\cdot\cdot\cdot, N-1$. Using the equation $T_{\mu_1\cdot\cdot\cdot\mu_p}^N(p_1,\cdot\cdot\cdot,p_{N-1},m_0,\cdot\cdot\cdot,m_{N-1})=\sum_{{i_1\cdot\cdot\cdot i_p=0}}^{{N-1}}T_{i_1\cdot\cdot\cdot i_p}^Np_{i_1\mu_1}\cdot\cdot\cdot p_{i_p\mu_p}$, the tensor integrals can be simplified, which is constructed from the external momenta $p_i$ and the symmetric coefficient functions $T_{i_1\cdot\cdot\cdot i_p}$. Therefore, the amplitude ${\cal A}_{\phi1}$ can be simplified as the invariant Passarino-Veltman integrals\cite{Passarino-Veltman}. Applying the high energy physics package FeynCalc\cite{FeynCalc}, we can also obtain the reduced result for the amplitude ${\cal A}_{\phi1}$.
\begin{eqnarray}
&&{\cal A}_{\phi1}=-\frac{ief_{\phi}m_{\phi}(3-4s_W^2)}{96\pi^2N_c[m_{\phi}^2-m_Z^2]*2s_Wc_W}
\bar{u}_{\mu}(p_4)A_3\{{/\!\!\!\varepsilon}^*(p)(2 a_{1L} C_{00}P_L+2a_{1R} C_{00}P_R)
\nonumber\\&&+p_4\cdot\varepsilon^*(p)[(m_{\mu}a_{1L}(2C_{22}+C_2)+m_ea_{1R}(2C_{12}+C_1)+m_{\chi_k^0}b_{1L}(C_0+2C_2))P_L
\nonumber\\&&+(m_{\mu}a_{1R}(2C_{22}+C_2)+m_e a_{1L}(2C_{12}+C_1)+m_{\chi_k^0}b_{1R}(C_0+2C_2))P_R]
\nonumber\\&&-p_3\cdot\varepsilon^*(p)[(m_{\mu}a_{1L}(2C_{12}+C_2)+m_ea_{1R}(2C_{11}+C_1)+m_{\chi_k^0}b_{1L}(C_0+2C_1))P_L
\nonumber\\&&+(m_{\mu}a_{1R}(2C_{12}+C_2)+m_e a_{1L}(2C_{11}+C_1)+m_{\chi_k^0}b_{1R}(C_0+2C_1))P_R]\}v_e(p_3),
\end{eqnarray}
where $a_{1L}=A_{2R}A_{1L}$, $a_{1R}=A_{2L}A_{1R}$,  $b_{1L}=A_{2L}A_{1L}$ and  $b_{1R}=A_{2R}A_{1R}$, $A_{1L}$, $A_{2L}$ and $A_{1R}$, $A_{2R}$ are the coefficients of left-handed and right-handed according to $A_1$ and $A_2$ respectively. In a similar way, the amplitudes of other penguin diagrams in FIG.1 can be obtained.

Then, taking process $\phi\rightarrow e^+\mu^-$ in FIG.2(1) as an example, the amplitude for box diagram is described as follows
\begin{eqnarray}
&&{\cal A}_{\phi_{b1}}=\langle e^+\mu^-|\bar{u}_{\mu}(p_4)
\int\frac{d^Dk}{(2\pi)^D}\frac{A_{b4}({/\!\!\!k}+{/\!\!\!p}_4+m_{\chi_i^0})A_{b1}}
{[(k+p_4)^2-m_{\chi_i^{\pm}}^2][(k+p_4-p_1)^2-m_{\tilde{U}}^2]}u_s(p_1)
\nonumber\\&&\hspace{1.2cm}\times\bar{v}_s(p_2)\frac{A_{b2}({/\!\!\!k}-{/\!\!\!p}_3+m_{\chi_j^{\pm}})A_{b3}}
{[(k-p_3)^2-m_{\chi_j^{\pm}}^2][k^2-m_{\tilde{\nu}}^2]}v_e(p_3)|\phi(p)\rangle,
\end{eqnarray}
where,
\begin{eqnarray}
&&A_{b1}=[(-\frac{e}{s_W}Z_U^{Lm*}Z_+^{1i}+Y_u^LZ_U^{(L+3)m*}Z_+^{2i})P_L
-Y_d^MZ_U^{Lm*}Z_-^{2i*}P_R]K^{LM},
\nonumber\\&&A_{b2}=(A_{b1})_{M\rightarrow N,i\rightarrow j}^*,
\nonumber\\&&A_{b3}=-[(\frac{e}{s_w}Z_+^{1j}Z_{\nu}^{Jk*}+Y_{\nu}^{Jk}Z_+^{2j}Z_{\nu}^{(J+3)k*})P_L
+{Y_l^{J}Z_-^{2j*}Z_{\nu}^{Jk*}}P_R],
\nonumber\\&&A_{b4}=(A_{b3})_{J\rightarrow I,j\rightarrow i}^*.
\end{eqnarray}

First, we need to swap the position of the wavefunctions $u_s(p_1)$ and $v_e(p_3)$. The method is named as Fierz Rearrangement, and the corresponding transformation rules and characters can be learnt from reference\cite{Fierz1,Fierz2,Fierz3}. After that, we can simplify amplitude ${\cal A}_{\phi_{b1}}$ by light-cone wavefunction and one-loop tensor integrals. The reduced results of amplitude ${\cal A}_{\phi_{b1}}$ can be written as:
\begin{eqnarray}
&&{\cal A}_{\phi_{b1}}=\frac{if_{\phi}}{128\pi^2N_c}\bar{u}_{\mu}(p_4)\{m_{\phi}{/\!\!\!\varepsilon}^*(p)[(a_{1L}'C_0
+(m_{\tilde{\nu}}^2a_{1L}'-2b_{1R}')D_0)P_L
\nonumber\\&&\hspace{1.4cm}+(a_{1R}'C_0+(m_{\tilde{\nu}}^2a_{1R}'-2b_{1L}')D_0)P_R]
-2[({/\!\!\!p}{/\!\!\!\varepsilon}^*(p)-{/\!\!\!\varepsilon}^*(p){/\!\!\!p})(c_{1L}'+c_{1R}')
\nonumber\\&&\hspace{1.4cm}+\sigma_{\mu\nu}\epsilon^{\mu\nu p\varepsilon^*(p)}(-c_{1L}'+c_{1R}')]D_0(P_L+P_R)\} v_e(p_3)
\end{eqnarray}
Here, $a_{1L}'=A_{b1L}A_{b2R}A_{b3L}A_{b4R}$, $b_{1L}'=A_{b1L}A_{b2R}(A_{b4R}m_{\mu}+A_{b4L}m_{\chi_i^{\pm}})(A_{b3L}m_{e}+A_{b3R}m_{\chi_j^{\pm}})$, $c_{1L}'=A_{b1L}A_{b2L}(A_{b4R}m_{\mu}+A_{b4L}m_{\chi_i^{\pm}})(A_{b3R}m_{e}+A_{b3L}m_{\chi_j^{\pm}})$,
$a_{1R}'=a_{1L}'|_{L\leftrightarrow R}$, $b_{1R}'=b_{1L}'|_{L\leftrightarrow R}$, $c_{1R}'=c_{1L}'|_{L\leftrightarrow R}$. $A_{b1L},\;A_{b2L},\;A_{b3L},\;A_{b4L}$ and $A_{b1R},\;A_{b2R},\;A_{b3R},\;A_{b4R}$ are the coupling coefficients of left-handed and right-handed corresponding to $A_{b1},\; A_{b2},\; A_{b3},\; A_{b4}$ respectively. The amplitudes of box diagrams in FIG.2 can be calculated by the same way.

The branching ratios for processes $V\rightarrow l_i^{\pm}l_j^{\mp}$ can be deduced as:
\begin{eqnarray}
&&Br\left(V\rightarrow l_i^{\pm}l_j^{\mp}\right)=\frac{\sqrt{[m_V^2-(m_{l_i}+m_{l_j})^2][m_V^2-(m_{l_i}-m_{l_j})^2]}}{16\pi m_V^3\Gamma_V}\times\sum_{\xi}A_{V_{\xi}}A_{V_{\xi}}^*,
\end{eqnarray}
where $\Gamma_V$ represents the total decay width of meson $V$ (with $V\in\{\phi, J/\Psi, \Upsilon, \rho^0, \omega \}$ and $l_i, l_j\in\{e, \mu, \tau \}$ ).  We chose $\Gamma_{\phi}\simeq4.2\times10^{-3}$ GeV, $\Gamma_{J/\Psi}\simeq0.093\times10^{-3}$ GeV, $\Gamma_{\Upsilon}\simeq0.054\times10^{-3}$ GeV, $\Gamma_{\rho^0}\simeq0.149$ GeV, $\Gamma_{\omega}\simeq8.49\times10^{-3}$ GeV\cite{PDG}. $A_{V_{\xi}}$ are the amplitudes corresponding to FIG.1 and FIG.2. Summation formula $\sum_{\lambda=\pm1,0}\varepsilon_{\lambda}^{\mu}(p)\varepsilon_{\lambda}^{*\nu}(p)=-g^{\mu\nu}+\frac{p^{\mu}p^{\nu}}{m_V^2}$ can be used to simplify $\sum_{\xi}A_{V_{\xi}}A_{V_{\xi}}^*$.
\section{Numerical Results}
In the numerical analysis, we consider the experimental constrains from the light neutral Higgs mass $m_{_{h^0}}\simeq125\;{\rm GeV}$\cite{LHC1,LHC2,LHC3} and the neutrino experiment data\cite{neutrino1,neutrino4,neutrino5,neutrino data1,neutrino data2}
\begin{eqnarray}
&&\sin^2\theta_{13}= (2.19\pm 0.12)\times10^{-2},~~\sin^2\theta_{12} =0.304\pm0.014, ~~\sin^2\theta_{23}=0.51\pm0.05,
\nonumber\\&&\Delta m_{\odot}^2 =(7.53\pm 0.18)\times 10^{-5} {\rm eV}^2, ~~|\Delta m_{A}^2| =(2.44\pm0.06)\times 10^{-3} {\rm eV}^2.
\end{eqnarray}
In our previous works, $Br(\mu\rightarrow e\gamma)<5.7\times10^{-13}$, $Br(\mu\rightarrow 3e)<1.0\times10^{-12}$ and $Br(Z\rightarrow e\mu)<7.5\times10^{-7}$\cite{PDG, constraints1,Br} are strict constrains for our parameter space.

In this work, the meson masses are adopted as $m_{\phi}=1.019$ GeV, $m_{J/\Psi}=3.096$ GeV, $m_{\Upsilon}=9.460$ GeV, $m_{\rho^0}=0.775$ GeV and $m_{\omega}=0.782$ GeV. The decay constants for the corresponding mesons are shown as $f_{\phi}=0.231$ GeV, $f_{J/\Psi}=0.405$ GeV, $f_{\Upsilon}=0.715$ GeV, $f_{\rho^0}=0.209$ GeV and $f_{\omega}=0.195$ GeV. Furthermore, some other parameters we used are shown as follows\cite{PDG,parameter}:
\begin{eqnarray}
 &&m_e=0.51\times10^{-3}{\rm GeV},~m_{\mu}=0.105{\rm GeV},~m_{\tau}=1.777{\rm GeV},L_4={3\over2},\nonumber\\&&m_u=2.2\times10^{-3}{\rm GeV},~m_c=1.27{\rm GeV},~m_t=173.2{\rm GeV},B_4={3\over2},\nonumber\\&&m_d=4.7\times10^{-3}{\rm GeV},~m_s=0.096{\rm GeV},~m_b=4.18{\rm GeV},~\lambda_{N^c}=1,\nonumber\\&&~m_Z=91.1876{\rm GeV},
m_W=80.385{\rm GeV},~\alpha(m_Z)=1/128,~s_W^2(m_Z)=0.23,
\nonumber\\&&(Y_{\nu})_{11}=1.3031*10^{-6},~(Y_{\nu})_{12}=9.0884*10^{-8},~(Y_{\nu})_{13}=6.9408*10^{-8},
\nonumber\\&&(Y_{\nu})_{22}=1.6002*10^{-6},~(Y_{\nu})_{23}=3.4872*10^{-7},~(Y_{\nu})_{33}=1.7208*10^{-6}.
\end{eqnarray}

We assume $(A_l)_{ii}=-2$ TeV, $(A_{N^c})_{ii}=(A_N)_{ii}=500$ GeV, $(A'_l)_{ii}=A'_L$, $(A_u)_{ii}=1$ TeV, $(A'_u)_{ii}=0.8$ TeV, $(A_d)_{ii}=1$ TeV, $(A'_d)_{ii}=1$ TeV, $\mu=0.7$ TeV, $\mu_L=0.5$ TeV, $(m_{\tilde{N}^c})_{ii}=1$ TeV and $(m_{\tilde{Q}})_{ii}=(m_{\tilde{U}})_{ii}=(m_{\tilde{D}})_{ii}=2$ TeV, here $i=1,2,3$. $\tan{\beta_L}=\bar{v}_L/v_L$ and $V_{L_t}=\sqrt{\bar{v}_L^2+v_L^2}$. The diagonal entries of matrices $m_{\tilde{L}}^2$ and $m_{\tilde{R}}^2$ are supposed as $(m_{\tilde{L}}^2)_{ii}=(m_{\tilde{R}}^2)_{ii}=S_m^2$ and non-diagonal terms $(m_{\tilde{L}}^2)_{ij}=(m_{\tilde{R}}^2)_{ij}=M_{L_f}^2$, with $i\neq j$ and $i,j=1,2,3$.
$m_1$ represents the gaugino mass in $U(1)$ and $m_2$ represents the gaugino mass in $SU(2)$. Generally, if we do not emphasize specially, the non-diagonal elements of the parameters are defined as zero.
\subsection{$V\rightarrow e^+\mu^-$}
At first, we discuss the CLFV decays of vector mesons $V\rightarrow e^+\mu^-$ with $V\in\{\phi, J/\Psi, \Upsilon, \rho^0, \omega \}$. The branching ratios for processes $\phi(J/\Psi)\rightarrow e^+\mu^-$ are strict. The corresponding experimental limits are $Br(\phi\rightarrow e^+\mu^-)\leq2.0\times 10^{-6}$ at $90\%$ confidence level and $Br(J/\Psi\rightarrow e^+\mu^-)\leq1.6\times 10^{-7}$ at $95\%$ confidence level.

As the diagonal elements of mass matrices $m_{\tilde{L}}^2$ and $m_{\tilde{R}}^2$, $S_m$ present in the mass matrices of slepton and sneutrino. And the CLFV processes can be influenced by slepton-neutralino, slepton-lepton neutralino and chargino-sneutrino contributions. At this subsection, the parameters are supposed as $m_1=0.5$ TeV, $m_2=0.5$ TeV, $g_L=0.1$, $\tan{\beta}_L=2$, $V_{L_t}=3$ TeV, $A'_L=0.3$ TeV, $M_L=1$ TeV and $M_{L_f}^2=2*10^5$ GeV$^2$. In FIG.3, we plot the branching ratios of decays $V\rightarrow e^+\mu^-$ varying with $S_m$. Here, different line corresponds to different decay process. We find that the results for processes $J/\Psi\rightarrow e^+\mu^-$ and $\Upsilon\rightarrow e^+\mu^-$ are around $10^{-9}\sim 10^{-10}$, the results of $\phi\rightarrow e^+\mu^-$ are around $10^{-10}\sim 10^{-13}$, and the results of $\rho^0\rightarrow e^+\mu^-$ and $\omega\rightarrow e^+\mu^-$ are around $10^{-11}\sim 10^{-14}$. These five lines all decrease quickly with the increasing $S_m$. Therefore, $S_m$ are very sensitive parameters to the numerical results.

\begin{figure}[h]
\centering
\includegraphics[width=10cm]{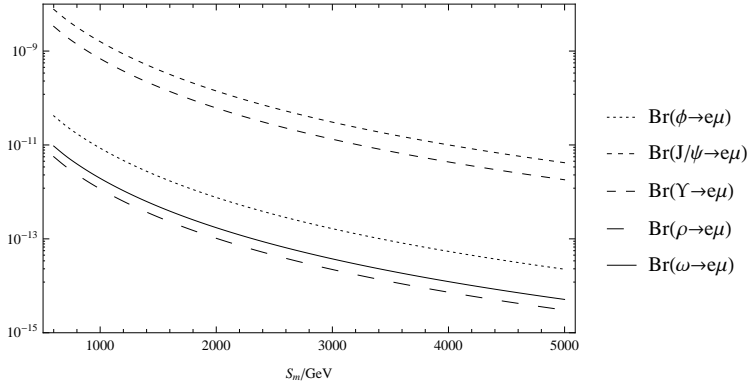}\\
\caption{The contributions to $V\rightarrow e^+\mu^-( V\in\{\phi, J/\Psi, \Upsilon, \rho^0, \omega \})$ varying with $S_m$ are respectively plotted by the five lines.} \label{fig3}
\end{figure}

The parameter $m_2$ not only presents in the mass matrix of neutralino, but also in the mass matrix of chargino. Therefore, $m_2$ affects the numerical results through the neutralino-slepton and chargino-sneutrino contributions. The branching ratios of $V\rightarrow e^+\mu^-$ varying with parameter $m_2$ are shown in FIG.4. Choosing $m_1=0.5$ TeV, $g_L=0.1$, $\tan{\beta}_L=2$, $V_{L_t}=3$ TeV, $A'_L=0.3$ TeV, $M_L=1$ TeV, $M_{L_f}^2=2*10^5$ GeV$^2$ and $S_m=1$ TeV, with the enlarging $m_2$, the branching ratios for each process all decrease. However, the results do not change very remarkable. Although $m_2$ is a sensitive parameter, the influence from $m_2$ is smaller than that from $S_m$.
\begin{figure}[h]
\centering
\includegraphics[width=10cm]{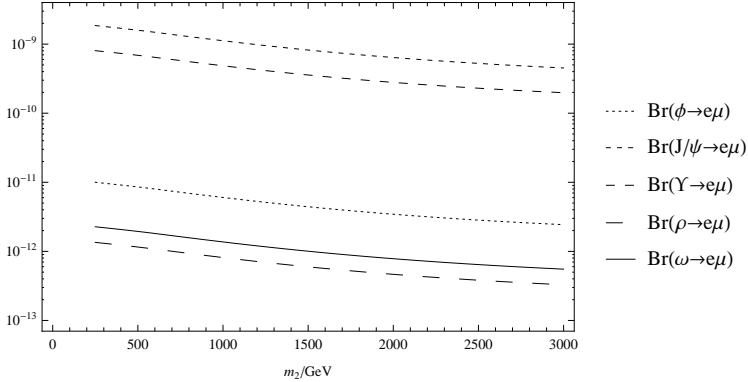}\\
\caption{ The decays $V\rightarrow e^+\mu^-(V\in\{\phi, J/\Psi, \Upsilon, \rho^0, \omega\})$ versus parameter $m_2$ are plotted by the five lines respectively.} \label{fig4}
\end{figure}

1. $J/\Psi(\Upsilon)\rightarrow e^+\mu^-$

Here, we study the decay processes $J/\Psi\rightarrow e^+\mu^-$ and $\Upsilon\rightarrow e^+\mu^-$. The neutralino mass matrix includes the parameter $m_1$, and the numerical results can be influenced by lepton-neutralino contributions. Supposing $g_L=0.1$, $\tan{\beta}_L=2$, $V_{L_t}=3$ TeV, $A'_L=0.3$ TeV, $M_L=1$ TeV, $M_{L_f}^2=2*10^5$ GeV$^2$, $S_m=1$ TeV and $m_2=0.5$ TeV, we give out the values of $Br(J/\Psi(\Upsilon)\rightarrow e^+\mu^-)$ versus $m_1$ in FIG.5. With the enlarging $m_1$, the numerical results decrease slightly. Therefore, $m_1$ is not a sensitive parameter for the CLFV decays.
\begin{figure}[h]
\centering
\includegraphics[width=10cm]{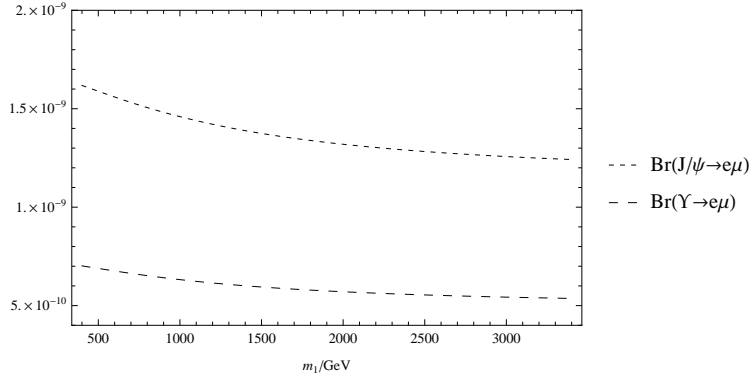}\\
\caption{The branching ratios of decays $J/\Psi(\Upsilon)\rightarrow e^+\mu^-$ change with the parameter $m_1$.} \label{fig5}
\end{figure}

Then, we study the processes with parameter $M_{L_f}$. As the non-diagonal elements of $m_{\tilde{L}}^2$ and $m_{\tilde{R}}^2$, $M_{L_f}$ lead
to strong mixing for slepton (sneutrino) of different generation. With $m_1=0.5$ TeV, $m_2=0.5$ TeV, $g_L=0.1$, $\tan{\beta}_L=2$, $V_{L_t}=3$ TeV, $A'_L=0.3$ TeV, $M_L=1$ TeV and $S_m=1$ TeV, the numerical results are plotted corresponding to parameter $M_{L_f}$ in FIG.6. When $M_{L_f}$ is zero, the branching ratios are almost zero; with enlarging $|M_{L_f}|$, the results for both processes increase quickly, which are around $0\sim10^{-8}$. And the figures are very symmetrical for both processes. Obviously, $M_{L_f}$ is a much sensitive parameter, and the effects from $M_{L_f}$ are very strong for these CLFV decays.
\begin{figure}[h]
\centering
\includegraphics[width=10cm]{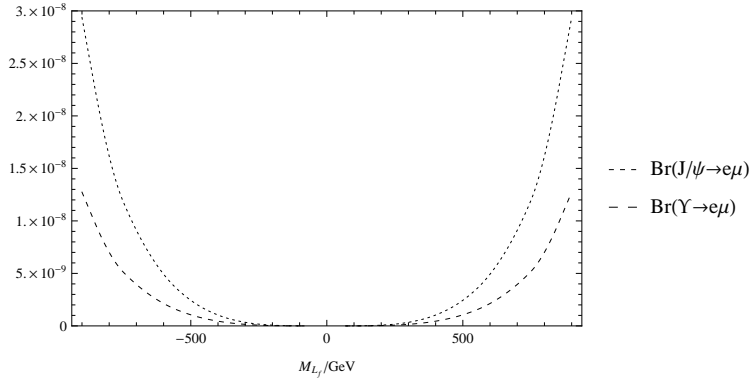}\\
\caption{The branching ratios for processes $J/\Psi(\Upsilon)\rightarrow e^+\mu^-$ change with the parameter $M_{L_f}$.} \label{fig6}
\end{figure}

2. $\phi(\rho^0,\omega)\rightarrow e^+\mu^-$

As a new parameters in the BLMSSM, $g_L$ and $\tan{\beta_L}$ have relation with the mass matrices of slepton, sneutrino and lepton neutralino. It is worth to consider the contributions from $g_L$ and $\tan{\beta_L}$. Based on the supposition $m_1=0.5$ TeV, $m_2=0.5$ TeV, $\tan{\beta}_L=2$, $V_{L_t}=3$ TeV, $A'_L=0.3$ TeV, $M_L=1$ TeV, $S_m=1$ TeV and $M_{L_f}^2=2*10^5$ GeV$^2$, the branching ratios for $\phi(\rho^0,\omega)\rightarrow e^+\mu^-$ are discussed with parameter $g_L$ in FIG.7. As $g_L>0.25$, the values for each process increase slightly. So the effects from $g_L$ are very small to the numerical results.
\begin{figure}[h]
\centering
\includegraphics[width=10cm]{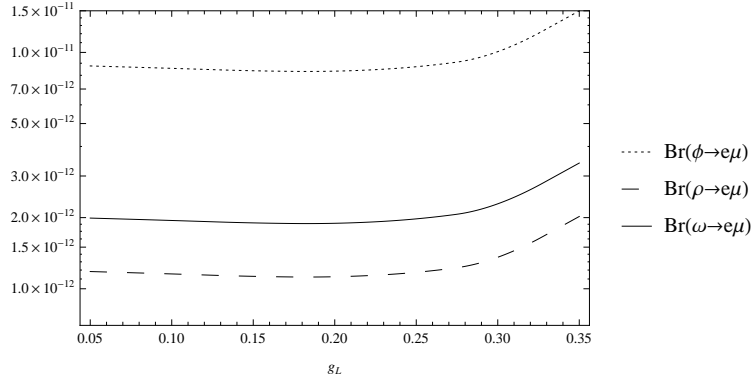}\\
\caption{The contributions to $Br(\phi(\rho^0,\omega)\rightarrow e^+\mu^-)$ varying with $g_L$ are plotted by the dotted line, dashed line and solid line respectively.} \label{fig7}
\end{figure}

The branching ratios versus $\tan{\beta_L}$ are studied. Here, $\tan{\beta_L}=\bar{v}_L/v_L$, $v_L$ and $\bar{v}_L$ are the nonzero VEVs of the $SU(2)_L$ singlets $\Phi_L$ and $\varphi_L$. In FIG.8, using $m_1=0.5$ TeV, $m_2=0.5$ TeV, $V_{L_t}=3$ TeV, $A'_L=0.3$ TeV, $M_L=1$ TeV, $S_m=1$ TeV, $M_{L_f}^2=2*10^5$ GeV$^2$ and $g_L=0.1$, we describe the numerical results for $\phi(\rho^0,\omega)\rightarrow e^+\mu^-$ with the parameter $\tan{\beta_L}$. These three lines almost keep the same values for each process with the enlarging $\tan{\beta_L}$. Obviously, the parameter $\tan{\beta_L}$ has tiny effects to our numerical analysis.
\begin{figure}[h]
\centering
\includegraphics[width=10cm]{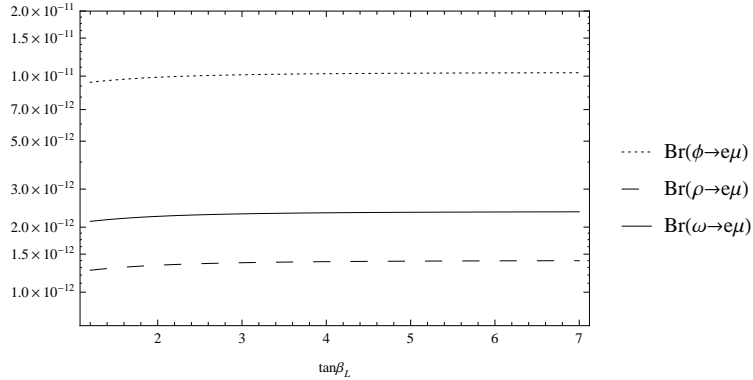}\\
\caption{The numerical results for $Br(\phi(\rho^0,\omega)\rightarrow e^+\mu^-)$ versus with $\tan{\beta_L}$ are plotted.} \label{fig8}
\end{figure}
\subsection{$V\rightarrow e^+\tau^-(\mu^+\tau^-)$}
We study the decays $V\rightarrow e^+\tau^-(\mu^+\tau^-)$ as follows, where $V\in\{J/\Psi,\Upsilon\}$. The experimental limits for decays $J/\Psi\rightarrow e^+\tau^-(\mu^+\tau^-)$ are $Br(J/\Psi\rightarrow e^+\tau^-)\leq8.3\times10^{-6}$ and $Br(J/\Psi\rightarrow \mu^+\tau^-)\leq2.0\times10^{-6}$, which are both at $90\%$ confidence level. The experimental limit for decay $\Upsilon\rightarrow \mu^+\tau^-$ is lower than $6.0\times10^{-6}$, which is at $95\%$ confidence level.

To study the processes $V\rightarrow e^+\tau^-(\mu^+\tau^-)$, the used parameters are $m_2=0.5$ TeV, $g_L=0.1$, $V_{L_t}=3$ TeV, $A'_L=0.3$ TeV, $M_L=1$ TeV, $S_m=1$ TeV, $M_{L_f}^2=2*10^5$ GeV$^2$ and $\tan{\beta}_L=2$. The numerical results versus $m_1$ are plotted in FIG.9. All of the processes decrease slightly with the enlarging $m_1$. When the decay processes have the same initial state, for example, $J/\Psi\rightarrow e^+\tau^-$ and $J/\Psi\rightarrow \mu^+\tau^-$, the numerical results for both processes are almost same. And the figure of $J/\Psi\rightarrow e^+\tau^-$ is under that of $J/\Psi\rightarrow \mu^+\tau^-$. Compared with the numerical results in FIG.5, the values of $Br(J/\Psi\rightarrow e^+\tau^-(\mu^+\tau^-))$ are lower than that of $Br(J/\Psi\rightarrow e^+\mu^-)$. The processes $\Upsilon\rightarrow l_i^{\pm}l_j^{\mp}$ have the same characters as $J/\Psi\rightarrow l_i^{\pm}l_j^{\mp}$.
\begin{figure}[h]
\centering
\includegraphics[width=10cm]{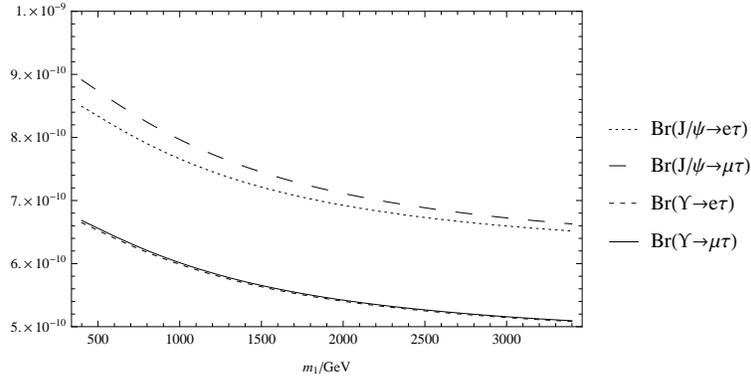}\\
\caption{The branching ratios for $Br(V\rightarrow e^+\tau^-(\mu^+\tau^-))(V\in\{J/\Psi,\Upsilon\})$ changing with $m_1$ are given out.} \label{fig9}
\end{figure}

Compared with MSSM, $V_{L_t}$ is also a new parameter, which presents in the mass matrices of slepton, sneutrino and lepton neutralino. With $V_{L_t}=\sqrt{v_L^2+\bar{v}_L^2}$, $m_{Z_L}=2g_LV_{L_t}$ is the mass of neutral $U(1)_L$ gauge boson $Z_L$. In FIG.10, the branching ratios for $V\rightarrow e^+\tau^-(\mu^+\tau^-)$ changing with $V_{L_t}$ are discussed, where $m_1=0.5$ TeV, $m_2=0.5$ TeV, $g_L=0.1$, $\tan{\beta}_L=2$, $A'_L=0.3$ TeV, $M_L=1$ TeV, $S_m=1$ TeV and $M_{L_f}^2=2*10^5$ GeV$^2$. When the values of $V_{L_t}$ change from 1 TeV to 4 TeV, each figure keeps an increscent variation trend. However, the numerical results all increase slowly, so the effects from $V_{L_t}$ are small.
\begin{figure}[h]
\centering
\includegraphics[width=10cm]{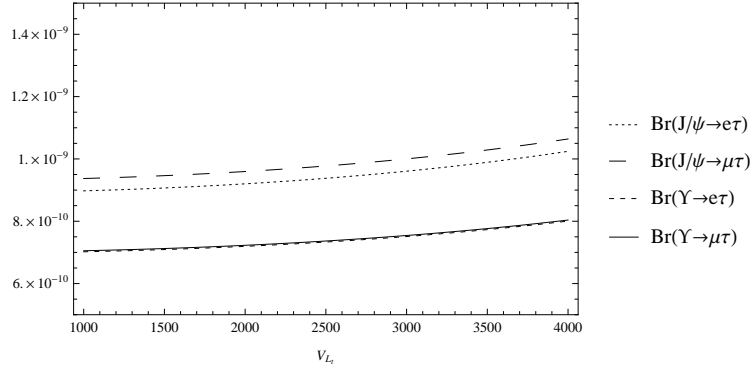}\\
\caption{For $Br(V\rightarrow e^+\tau^-(\mu^+\tau^-)),V\in\{J/\Psi,\Upsilon\}$, the results varying with $V_{L_t}$ are plotted.} \label{fig10}
\end{figure}

1. $J/\Psi(\Upsilon)\rightarrow \mu^+\tau^-$

The decays $J/\Psi\rightarrow e^+\tau^-$ and $J/\Psi\rightarrow \mu^+\tau^-$ almost possess the same variation trend, processes $\Upsilon\rightarrow e^+\tau^-$ and $\Upsilon\rightarrow \mu^+\tau^-$ also have the same character. Therefore, we just consider the decays $J/\Psi(\Upsilon)\rightarrow \mu^+\tau^-$ in this section. The slepton mass squared matrix includes the parameter $A'_L$, which is the non-diagonal element of this matrix. So the results can be affected by $A'_L$ through slepton mass and mixing. Here, $m_1=0.5$ TeV, $m_2=0.5$ TeV, $g_L=0.1$, $\tan{\beta}_L=2$, $V_{L_t}=3$ TeV, $M_L=1$ TeV, $S_m=1$ TeV and $M_{L_f}^2=2*10^5$ GeV$^2$, the branching ratios for $J/\Psi(\Upsilon)\rightarrow \mu^+\tau^-$ versus parameter $A'_L$ are plotted in FIG.11. With $A'_L$ changing from 0 to 5 TeV, both processes slightly increase. And the diagram for $J/\Psi\rightarrow \mu^+\tau^-$ is under that of $\Upsilon\rightarrow \mu^+\tau^-$. Obviously, the effects from $A'_L$ are not large to the numerical results.

\begin{figure}[h]
\centering
\includegraphics[width=10cm]{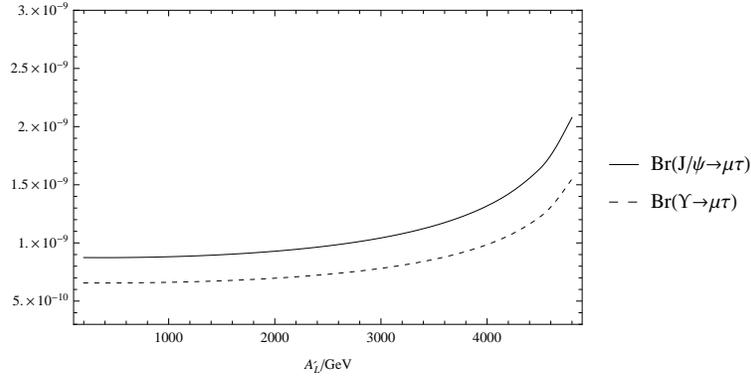}\\
\caption{The numerical results of $Br(J/\Psi(\Upsilon)\rightarrow \mu^+\tau^-)$ varying with $A'_L$ are plotted by the solid line and dotted line respectively.} \label{fig9}
\end{figure}

As a diagonal element of lepton neutralino mass matrix, $M_L$ is a new introduced parameter in the BLMSSM. To see how $M_L$ affects the branching ratios of $J/\Psi(\Upsilon)\rightarrow \mu^+\tau^-$, with $m_1=0.5$ TeV, $m_2=0.5$ TeV, $g_L=0.1$, $\tan{\beta}_L=2$, $V_{L_t}=3$ TeV, $S_m=1$ TeV, $M_{L_f}^2=2*10^5$ GeV$^2$ and $A'_L=0.3$ TeV, we give out the allowed numerical results varying with $M_L$. When $M_L$ changes from $0$ to $1$ TeV, the results for both processes decrease slightly; when $M_L>1$ TeV, the values for each process almost keep same. The numerical results indicate that the influence from $M_L$ is not obvious, and can be neglected.

\begin{figure}[h]
\centering
\includegraphics[width=10cm]{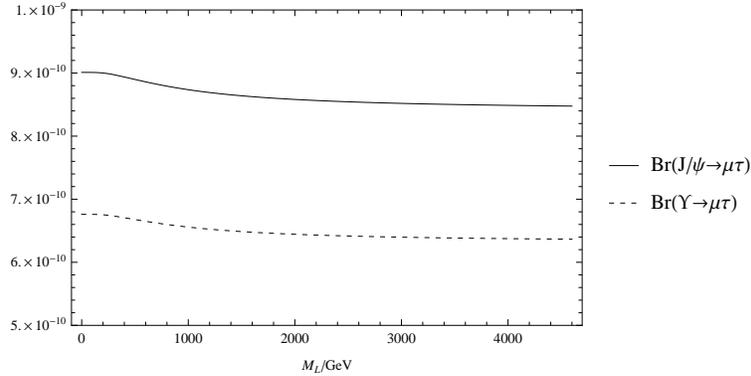}\\
\caption{With the parameter $M_L$, the numerical results for $Br(J/\Psi(\Upsilon)\rightarrow \mu^+\tau^-)$ are shown.} \label{fig9}
\end{figure}
\section{discussion and conclusion}
In the frame of BLMSSM, we study the CLFV decays for vector mesons $V\rightarrow l_i^{\pm}l_j^{\mp}$ with $V\in\{\phi, J/\Psi, \Upsilon, \rho^0, \omega \}$ and $l_i, l_j\in\{e, \mu, \tau \}$. In this model, new parameters and new contributions are considered to these CLFV processes. For example, with the new introduced  parameters $g_L$, $\tan{\beta_L}$ and $V_{L_t}$, lepton neutralino $\chi_L^0$ is discussed in our work, which give new type contributions through the lepton neutralino-slepton-lepton coupling. Furthermore, three heavy neutrinos and three new scalar neutrinos are also considered in the BLMSSM. The contributions from neutrinos can be neglected due to the tiny Yukawa $Y_\nu$. However, the new scalar neutrinos play very important roles, especially the non-diagonal elements $M_{L_f}$ in $(m_{\tilde{L}}^2)_{IJ}$, which lead to strong mixing for scalar neutrinos of different generation and enhance the lepton flavor violation.

Considering the numerical results discussed in the Section IV, various parameters affect the CLFV decays. $S_m$ and $M_{L_f}$ are the sensitive parameters, which are the diagonal and non-diagonal elements in matrices $m_{\tilde{L}}$ and $m_{\tilde{R}}$. The influence from $S_m$ and $M_{L_f}$ is very remarkable. After discussing the constraints from $l_j\to 3l_i$, $l_j\to l_i+\gamma$ and $\mu-e$ conversion, $Br(\phi\rightarrow e\mu)\sim10^{-11}, Br(J/\Psi(\Upsilon)\rightarrow e\mu)\sim10^{-9}$ and $Br(\rho^0(\omega)\rightarrow e\mu)\sim2\times10^{-12}$, the decay $J/\Psi(\Upsilon)\rightarrow e\mu$ is much easier than $\phi(\rho^0,\omega)\rightarrow e\mu$ to reach the experimental upper bounds. Similarly, $Br(J/\Psi\rightarrow e\tau(\mu\tau))$ are at the order of $(10^{-10}\sim10^{-9})$, and $Br(\Upsilon\rightarrow e\tau(\mu\tau))$ can reach $5\times10^{-10}$, which are very promising to be observed in the near future experiments.
\section{Acknowledgments}
Supported by the Major Project of
NNSFC (No. 11535002, No. 11605037, No. 11647120, No. 11275036),
the Natural Science Foundation of Hebei province with Grant
No. A2016201010 and No. A2016201069, and the Natural Science Fund of
Hebei University with Grants No. 2011JQ05 and No. 2012-
242, Hebei Key Lab of Optic-Electronic Information and
Materials, the midwest universities comprehensive strength
promotion project.
\section{Appendix A}
In this section, we give out the corresponding superfields presented in BLMSSM model, which are shown in TABLE I:
\begin{table}[h]
\caption{ \label{tab1}  Superfields in the BLMSSM.}
\begin{tabular*}{87.5mm}{|c|c|c|c|c|c|}
\hline
Superfields &$SU(3)_C$ &$SU(2)_L$ &$U(1)_Y$ &$U(1)_B$ &$U(1)_L$ \\
\hline
$\hat{Q}$&3 &2 &1/6 &1/3 &0 \\
\hline
$\hat{U}^c$&$\bar{3}$ &1 &-2/3 &-1/3 &0 \\
\hline
$\hat{D}^c$&$\bar{3}$ &1 &1/3 &-1/3 &0 \\
\hline
$\hat{L}$&1 &2 &-1/2 &0 &1 \\
\hline
$\hat{E}^c$&1 &1 &1 &0 &-1 \\
\hline
$\hat{N}^c$&1 &1 &0 &0 &-1 \\
\hline
$\hat{Q}_4$&3 &2 &1/6 &$B_4$ &0 \\
\hline
$\hat{U}_4^c$&$\bar{3}$ &1 &-2/3 &-$B_4$ &0 \\
\hline
$\hat{D}_4^c$&$\bar{3}$ &1 &1/3 &-$B_4$ &0 \\
\hline
$\hat{L}_4$&1 &2 &-1/2 &0 &$L_4$ \\
\hline
$\hat{E}_4^c$&1 &1 &1 &0 &-$L_4$ \\
\hline
$\hat{N}_4^c$&1 &1 &0 &0 &-$L_4$ \\
\hline
$\hat{Q}_5^c$&$\bar{3}$ &2 &-1/6 &-1-$B_4$ &0 \\
\hline
$\hat{U}_5$&3 &1 &2/3 &1+$B_4$ &0 \\
\hline
$\hat{D}_5$&3 &1 &-1/3 &1+$B_4$ &0 \\
\hline
$\hat{L}_5^c$&1 &2 &1/2 &0 &-3-$L_4$ \\
\hline
$\hat{E}_5$&1 &1 &-1 &0 &3+$L_4$ \\
\hline
$\hat{N}_5$&1 &1 &0 &0 &3+$L_4$ \\
\hline
$\hat{H}_u$&1 &2 &1/2 &0 &0 \\
\hline
$\hat{H}^d$&1 &2 &-1/2 &0 &0 \\
\hline
$\hat{\Phi}_B$&1 &1 &0 &1 &0 \\
\hline
$\hat{\varphi}_B$&1 &1 &0 &-1 &0 \\
\hline
$\hat{\Phi}_L$&1 &1 &0 &0 &-2 \\
\hline
$\hat{\varphi}_L$&1 &1 &0 &0 &2 \\
\hline
$\hat{X}$&1 &1 &0 &2/3+$B_4$ &0 \\
\hline
$\hat{X}'$&1 &1 &0 &-2/3-$B_4$ &0 \\
\hline
\end{tabular*}%
\end{table}

 \end{document}